\documentclass[aps,pra,showpacs,amsmath,amssymb,lengthcheck]{revtex4}
\usepackage{epsfig}

\begin{document}

\title{Interface Tension of Bose-Einstein Condensates}

\author{{\small Bert Van Schaeybroeck}}
\address{{\small\it Instituut voor Theoretische Fysica,}\\{\small\it Katholieke Universiteit Leuven}, {\small\it Celestijnenlaan 200 D, B-3001 Leuven, Belgium.}}

\date{\small\it \today}

\begin{abstract}
Motivated by recent observations of phase-segregated binary
Bose-Einstein condensates, we propose a method to calculate the
excess energy due to the interface tension of a trapped
configuration. By this method one should be able to numerically
reproduce the experimental data by means of a simple Thomas-Fermi
approximation, combined with interface excess terms and the
Laplace equation. Using the Gross-Pitaevskii theory, we find
expressions for the interface excesses which are accurate in a
very broad range of the interspecies and intraspecies interaction
parameters. We also present finite-temperature corrections to the
interface tension which, aside from the regime of weak
segregation, turn out to be small.
\end{abstract}

\pacs{03.75.Hh, 67.60.Bc, 67.85.-d, 67.85.Bc, 67.85.Fg}

\maketitle
\section{Introduction}
Soon after the first experimental realization of Bose-Einstein
condensation (BEC) in dilute gases, binary BEC mixtures were
realized~\cite{modugno,
miesner,myatt,stamper,hall,matthews,stenger,thalhammer,papp,papp2}.
These experiments precipitated a strong theoretical interest, the
origin of which is the fact that the multicomponent BEC is not a
trivial extension of the single-component
BEC~\cite{ho,graham,pu,esry,esry2,ohberg,alexandrov,timmermans,law,bashkin,burke,busch,riboli,jezek,mazets,ao,ao2,svidzinsky,barankov,chui}.
Indeed, novel and fundamentally different physics arise on both
microscopic and macroscopic levels. Phenomena studied for binary
mixtures so far include the quantum tunnelling of spin
domains~\cite{stamper,miesner}, spin-relaxation
processes~\cite{miesner}, vortex configurations~\cite{matthews},
formation of Feshbach molecules~\cite{papp} and collective
oscillations~\cite{modugno}. Experiments with phase-segregated BEC
mixtures were established already in 1998~\cite{stenger,hall};
this triggered many theorists to focus on the observed
weakly-segregated phases whereby the importance of the interface
physics was highlighted several
times~\cite{ao,ao2,chui,alexandrov,svidzinsky,mazets,timmermans,barankov,riboli,jezek}.

Also surfaces of \textit{single}-component BEC gases have gained
much attention. Studies are performed on vortices, surface modes
and tunnelling phenomena at the border of a trapped
single-component BEC~\cite{lundh,dalfovo,khawaja,fetter2} and on
BEC diffraction and van der Waals forces using an optical mirror
or evanescent-wave prism~\cite{colombe,rychtarik,landragin}.
Moreover a combination of the interface physics of binary BEC
gases and the surface physics near hard walls led to the
theoretical prediction of anomalous wetting phase
transitions~\cite{indekeu}.

The clearest observation so far of phase segregation of BEC
mixtures was reported by Papp et al.~(see Ref.~\onlinecite{papp2})
who used a mixture of $^{85}$Rb and $^{87}$Rb particles. By
changing the particle numbers of both species and the intraspecies
interactions of the $^{85}$Rb particles by use of a Feshbach
resonance, many topologically distinct states were encountered. In
the phase-segregated regime most trap configurations contained BEC
droplets. Papp et al.~stated that a detailed theoretical
understanding of the observed droplet formation is lacking. As an
essential first step towards this understanding we present here
expressions for the interface tension. Moreover, we also formulate
a method by which all trap configurations can be straightforwardly
studied by the use of a local density approximation or
Thomas-Fermi approximation. At first sight, numerically
reproducing the experimentally observed (meta)stable states
requires solving the coupled Gross-Pitaevskii (GP) equations for
the full three-dimensional system; in such case, a high accuracy
is indispensable to capture the important energy contributions
near the two-phase boundaries which strongly affect the ground
state topology.

In this work, we argue that it is not necessary to solve the full
GP system in order to recover numerically the experimental
configurations; instead, one can find the ground state, that is,
minimize the total energy of the trapped cloud, by use of the
Thomas-Fermi approximation and the expressions for the interface
tension which we provide here. This allows an omission of the
quantum pressure term which strongly complicates the numerical
work. The analytical calculation of the interface tension is
nontrivial as it involves solving the binary GP equations
(i.e.~two coupled nonlinear second-order differential equations).
Estimates and analytical expressions were already given in
Refs.~\onlinecite{timmermans}-\onlinecite{barankov} and were
relevant to the early experiments on weakly-segregated mixtures
but irrelevant to the interfaces as observed in
Ref.~\onlinecite{papp2} since there segregation is not weak. By
developing expansions, we succeed in obtaining analytical
expressions for the interface tension which are accurate for
almost all parameter regimes.

As a main result we find that, even though the interface tension
is not constant throughout the trap, the total excess energy
$\Omega_\mathcal{A}$ can be reduced to a simple integral of the
trapping potential along the interface area $\mathcal{A}$:
\begin{align}\label{totalexcess}
\Omega_\mathcal{A}&=\frac{\sqrt{2m_{_1}} }{4\pi\hslash
a_{_{11}}}\mathcal{F}(\xi_{_2}/\xi_{_1},K)
\int_{_\mathcal{A}}\text{d}\mathbf{r}\,[\mu_{_1}-U_{_1}(\mathbf{r})]^{3/2},
\end{align}
where $a_{_{11}}$, $m_{_1}$, $\mu_{_1}$ and $U_{_1}$ are the
intraspecies scattering length, the particle mass, the chemical
potential and the applied trapping potential of phase $1$. The
parameters $K$ and $\xi_{_2}/\xi_{_1}$ can be expressed in terms
of the atomic masses and the interspecies and intraspecies
scattering lengths:
\begin{align}\label{kenxi}
K=\frac{m_{_1}+m_{_2}}{2\sqrt{m_{_1}m_{_2}}}\frac{a_{_{12}}}{\sqrt{a_{_{11}}a_
{_{22}}}}\text{
 and  }\xi_{_2}/\xi_{_1}=\sqrt[4]{\frac{m_{_1}a_{_{11}}}{m_{_2}a_{_{22}}}}.
\end{align}
For a broad range of values of $\xi_{_2}/\xi_{_1}$ and $K$, the
function $\mathcal{F}$ can be written as:
\begin{align}\label{strongrepultsion}
&\mathcal{F}(\xi_{_2}/\xi_{_1},K)=\frac{\sqrt{2}}{3}(1+\xi_{_2}/\xi_{_1})
-\frac{0.514\sqrt{\xi_{_2}/\xi_{_1}}}{K^{1/4}}\\
&-\sqrt{\xi_{_2}/\xi_{_1}}(\xi_{_2}/\xi_{_1}+\xi_{_1}/\xi_{_2})\left(\frac{0.055}{K^{3/4}}+\frac{0.067}{K^{5/4}}\right)+\ldots\nonumber.
\end{align}

This work is structured as follows. We start off in
Sect.~\ref{sec_bosonboson} by introducing the GP formalism for
binary BECs. After defining the excess energy in a homogeneous and
in a trapped system in Sect.~\ref{sec_GCE}, we find analytical
expressions for the interface tension in Sect.~\ref{sec_inttens}.
We estimate the finite-temperature corrections to the interface
tension in Sect.~\ref{sec_tempdep}. In Sect.~\ref{sec_exp} we
discuss the experimental relevance of our findings.

\section{Binary BEC System}~\label{sec_bosonboson}
Equilibrium states of a mixture of two dilute BEC gases with order
parameters $\Psi_{_1}$ and $\Psi_{_2}$ and chemical potentials
$\mu_{_1}$ and $\mu_{_2}$, are well modelled using the grand
potential:
\begin{widetext}
\begin{align}\label{vrije1}
\Omega(\mu_{_1},\mu_{_2},V)=&\sum_{i=1,2}\left(\int_{_V}{\text{d}\mathbf{r}\,
\Psi_{_i}^*(\mathbf{r})\left[-\frac{\hslash^{2}}{2m_{_i}}
\boldsymbol{\nabla}^{2}-\mu_ {_i}+U_{_i}(\mathbf{r})\right]
\Psi_{_i}(\mathbf{r})}+\frac{G_{_{ii}}}{2}|\Psi_{_i}(\mathbf{r})|^{4}\right)
+G_{_{12}}\int_{_V}{\text{d}\mathbf{r}\,|\Psi_{_1}(\mathbf{r})|^{2}
|\Psi_{_2}(\mathbf{r})|^{2}},
\end{align}
\end{widetext}
where $G_{_{ij}}=2\pi
\hslash^{2}a_{_{ij}}(m_{_i}^{-1}+m_{_j}^{-1})$ are the coupling
constants and $a_{_{ij}}$ are the s-wave scattering lengths
(henceforth $i=1,\,2$). In view of the derivation of the interface
tension, we continue here by assuming vanishing external
potentials $U_{_1}=U_{_2}=0$; we reintroduce the external
potentials when discussing the surface excess of a trapped system.
In the absence of particle flow one can choose the order
parameters to be real valued such that the equilibrium pressures
for the \textit{pure} states are:
\begin{align}\label{drukken}
P_{_i}=\frac{\mu^{2}_{_i}}{2G_{_{ii}}}.
\end{align}
In order to study phase-segregated states of the two coexisting
condensates, we introduce the parameter
\begin{align}\label{K}
K\equiv\frac{G_{_{12}}}{\sqrt{G_{_{11}}G_{_{22}}}},
\end{align}
which quantifies the interspecies couplings relative to the
average of the intraspecies couplings. It is well-known that when
$K$ exceeds one, the species become immiscible and only pure
phases can exist~\cite{ao} while phase mixing occurs when $K<1$.
Along the two-phase interface, coexistence is ensured by the
condition:
\begin{align}\label{coexistence}
\frac{\mu_{_1}^2}{2G_{_{11}}}=\frac{\mu_{_2}^2}{2G_{_{22}}}.
\end{align}
Henceforth we assume that $K>1$ and that the chemical potentials
are such that the bulk pressures are $P_{_1}=P_{_2}\equiv P$. For
calculational convenience, we further rescale the order parameters
$\Psi_{_1}$ and $\Psi_{_2}$ and define the dimensionless wave
functions $\psi_{_1}$ and $\psi_{_2}$:
\begin{subequations}\label{relativedens}
\begin{align}
\Psi_{_i}\equiv \psi_{_i}\sqrt{\frac{\mu_{_i}}{G_{_{ii}}}}.
\end{align}
\end{subequations}
The quantum nature of the system results in a zero-point motion;
this determines the typical length scale for density modulations
at boundaries, impurities, vortices or solitons. For the pure
phases, the resultant length is the healing length $\xi_{_i}$
which is defined as:
\begin{align}\label{cohlength}
\xi_{_i}\equiv\frac{\hslash}{\sqrt{2m_{_i}\mu_{_i}}}.
\end{align}
Without loss of generality, we choose the phases such that
$\xi_{_2}/\xi_{_1}\leq 1$. As coexistence must occur along the
interface, $K$ and $\xi_{_2}/\xi_{_1}$ can be expressed in terms
of the atomic masses and the scattering lengths as given in
Eq.~\eqref{kenxi}. Finally, after rescaling the space coordinate
to the dimensionless variable
$\widetilde{\mathbf{r}}\equiv\mathbf{r}/\xi_{_1}$, the reduced
time-independent Gross-Pitaevskii (GP)
Eqs.~are~\cite{pita,pethick}:
\begin{subequations}\label{GP}
\begin{align}
\boldsymbol{\nabla}^2\psi_{_1}=&
-\psi_{_1}+\psi_{_1}^{3}+K\psi_{_1}\psi_{_2}^{2},\label{GPa}\\
[\xi_{_2}/\xi_{_1}]^2\boldsymbol{\nabla}^2\psi_{_2}=&
-\psi_{_2}+\psi_{_2}^{3}+K\psi_{_2}\psi_{_1}^{2}.\label{GPb}
\end{align}
\end{subequations}
Having established the equations of motion, we continue now by
defining the interface tension.

\section{Definition of Interface Tension and interface Excess}\label{sec_GCE}
In order to calculate the interface tension, consider two BEC
components in an infinitely large system with translational
symmetry in the $x$-$y$ direction. The presence of phase $1$ at
$z\rightarrow\infty$ and of phase $2$ at $z\rightarrow-\infty$
imply the following boundary conditions:
\begin{subequations}\label{boundconds}
\begin{align}
\psi_{_1}(z&=-\infty)=\psi_{_2}(z=\infty)=0,\\
\psi_{_1}(z&=\infty)=\psi_{_2}(z=-\infty)=1.
\end{align}
\end{subequations}
Under these conditions, the first integral of the reduced GP
Eqs.~\eqref{GPa} and~\eqref{GPb} is:
\begin{align}\label{conservation}
\dot{\psi}_{_1}^{2}+[\xi_{_2}/\xi_{_1}]^2\dot{\psi}_{_2}^{2}-K
\psi_{_1}^{2}\psi_{_2}^{2}+\sum_{_{i=1,2}}\left(
\psi_{_i}^{2}-\frac{\psi_{_i}^{4}}{2}\right)=\frac{1}{2},
\end{align}
where the dot denotes the derivative with respect to
$\widetilde{z}$. Since we work at fixed chemical potentials the
interface tension is the excess grand potential per unit area;
this excess is uniquely determined by subtraction of the total
grand potential of a volume $V$ containing a pure phase, from the
grand potential $\Omega$, that is, $\Omega+PV$. By use of
Eqs.~\eqref{GP} and \eqref{conservation} the \textit{interface
tension} can be written as:
\begin{align}\label{interfacetension}
\gamma_{_{12}}=&\,4P\xi_{_1}\int_{_{-\infty}}^{_{+\infty}}\text{d}\widetilde{z}\,\left[
\dot{\psi^{2}_{_1}}(\widetilde{z})+[\xi_{_2}/\xi_{_1}]^2\dot{\psi^{2}_{_2}}(\widetilde{z})\right]\nonumber\\
\equiv& 4P\xi_{_1}\,\mathcal{F}(\xi_{_2}/\xi_{_1},K),
\end{align}
The interface tension in the grand canonical ensemble is exactly
four times the interface tension in the canonical ensemble as
introduced by Ao and Chui~\cite{ao}~\footnote{Moreover the
interface tension in the canonical ensemble $\gamma_{_{AC}}$ as
introduced by Ao and Chui can be related to $\gamma_{_{12}}$ by a
surface analogue of a Legendre transformation
$\gamma_{_{AC}}(N_{_1},N_{_2},V)=\gamma_{_{12}}(\mu_{_1},\mu_{_2},V)-(PV-\mu_{_1}N_{_1}-\mu_{_2}N_{_2})/\mathcal{A}$
where $N_{_i}=(\partial[\gamma_{_{12}}-PV]/\partial\mu_{_i})_{_V}$
for $i=1,\,2$.}. Note also that, as opposed to the grand canonical
ensemble, the definition of the interface tension in the canonical
ensemble is not unique~\cite{fetter}. In
Eq.~\eqref{interfacetension}, we define the dimensionless function
$\mathcal{F}$ which, as is also the case for the normalized
profiles $\psi_{_i}$, only depends on $\xi_{_2}/\xi_{_1}$ and $K$
since the wave functions are fully determined by the GP
Eqs.~\eqref{GP} and the boundary conditions~\eqref{boundconds}. It
is clear from Eq.~\eqref{interfacetension} that the excess energy
is positive and due to a bending of the normalized wave functions
$\psi_{_1}$ and $\psi_{_2}$.

One can now straightforwardly generalize the definition of the
interface tension to the total excess energy of a trapped system.
Therefore we use a \textit{local approximation} for the interface
tension: we assume the characteristic lengths over which the
trapping potentials $U_{_i}(\mathbf{r})$ (for $i=1,\,2$) vary to
be large as compared to the interface thickness. At each point
$\mathbf{r}$ along the interface, Eq.~\eqref{interfacetension}
then gives the expression for the interface tension
$\gamma_{_{12}}(\mathbf{r})$ for condensates at effective chemical
potentials $\mu_{_i}-U_{_i}(\mathbf{r})$ ($i=1,\,2$). The total
excess energy $\Omega_\mathcal{A}$ is then obtained by integration
over the interface area $\mathcal{A}$ i.e.
$\Omega_\mathcal{A}=\int_{_\mathcal{A}}\text{d}\mathbf{r}\,\gamma_{_{12}}(\mathbf{r})$.
The function $\mathcal{F}(\xi_{_2}/\xi_{_1},K)$ can be put outside
this integral as it is position-independent; indeed, both $K$ and
$\xi_{_2}/\xi_{_1}$ are uniquely determined by the scattering
lengths and the particle masses (see Eq.~\eqref{kenxi}). This
remarkable property implies that, first of all, along the entire
interface in a trap, the interface profiles $\psi_{_1}$ and
$\psi_{_2}$ are unchanged (determined by Eqs.~\eqref{GP} and
\eqref{boundconds} only). This is an exceptional feature of BEC
interfaces~\footnote{For ``general'' mixtures, the normalized
density profiles across the interface will depend on the position
in the trap. For example, consider mixtures for which the
interspecies interactions are via a contact potential and for
which the pressures of the pure phases $P_{_i}$ and their
densities relate as $P_{_i}\propto n_{_i}^{\nu_{_i}}$ (with
$i=1,2$), and for which GP-like equations are valid. In that case,
the normalized density profiles do depend on position unless
$\nu_{_1}=\nu_{_2}=2$, as valid for BEC mixtures. The dependence
is for instance present for boson-fermion interfaces at zero
temperature.}. Secondly, it implies that in a trap the local
interface tension $\gamma_{_{12}}(\mathbf{r})$ depends on position
only through the term $P\xi_{_1}$ as can be seen from
Eq.~\eqref{interfacetension}. The \textit{total excess grand
potential} is then:
\begin{align}\label{totalexcess2}
\Omega_\mathcal{A}&=\,4\,\mathcal{F}(\xi_{_2}/\xi_{_1},K)\int_{_\mathcal{A}}\text{d}\mathbf{r}\,P(\mathbf{r})\xi_{_1}(\mathbf{r}),
\end{align}
which, using the effective chemical potentials
$\mu_{_i}-U_{_i}(\mathbf{r})$ in Eqs.~\eqref{drukken} and
\eqref{cohlength}, brings us to Eq.~\eqref{totalexcess} wherein we
factorize the total excess energy as a product of two distinct
terms: the first is a function of the microscopic gas parameters
whereas the last is a position integral of the external potential
along the interface area $\mathcal{A}$. The calculation of the
function $\mathcal{F}$ constitutes the subject of next section.

Expression~\eqref{totalexcess} for the total excess energy is our
main result. Together with a Thomas-Fermi (TF) approximation in
bulk, it forms the core of a simple method to calculate the total
energy of the trapped system and therefore to minimize it. The TF
approach neglects the energy contributions arising from density
gradients and thus gives rise to sharp interfaces, the locus of
which is normally taken where two-phase coexistence is satisfied.
The latter, however, is invalid when the interface is curved as
appears for example for droplets. Instead, a difference in
pressures appears along the two-phase boundary as expressed by the
Laplace equation:
\begin{align}\label{laplace}
P_{_1}-P_{_2}=\gamma_{_{12}}\left(1/R_{_1}+1/R_{_2}\right),
\end{align}
with $R_{_1}$ and $R_{_2}$ the principal radii of curvature of the
interface. It is clear that the difference in pressures along the
interface is largest for small droplets. Within a first approach,
for large $R_{_1}$ and $R_{_2}$, one can take the $\gamma_{_{12}}$
in Eq.~\eqref{laplace} to be the interface tension for a flat
interface, calculated by assuming equal pressures on opposite
sides of the interface.

Finally, note that, pertaining to the interface tension in a
\textit{trapped} system, it is essential to work at fixed chemical
potentials instead of particle numbers and use
definition~\eqref{interfacetension} as opposed to the interface
tension in the canonical ensemble~\cite{ao,fetter}.

\section{Calculation of Interface Tension}\label{sec_inttens}
\begin{figure}
\begin{center}
    \epsfig{figure=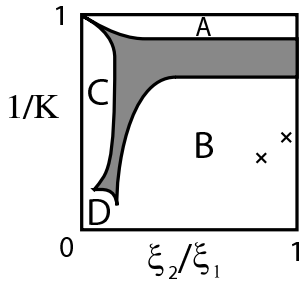,angle=0,width=110pt}
        \caption{Regions of validity of the approximations presented in Sects.~\ref{sec_inttens} A-D in the
        space of parameters $1/K$ and $\xi_{_2}/\xi_{_1}$. In the
        shaded region, the approximations have a low accuracy; the boundaries are drawn on a qualitative basis. The
        crosses denote values corresponding to the experiments
        of Ref.~\cite{papp2}.
        \label{fig1}
        }
  \end{center}
\end{figure}

In the following four subsections A-D, we derive expressions for
$\mathcal{F}(\xi_{_2}/\xi_{_1},K)=\gamma_{_{12}}/(4P\xi_{_1})$ as
expansions for different regimes of $\xi_{_2}/\xi_{_1}$ and $K$.
The wave functions $\psi_{_1}$ and $\psi_{_2}$ must satisfy the
boundary conditions~\eqref{boundconds}. The regions of validity of
the expansions are outlined (qualitatively) in Fig.~\ref{fig1}
where A-D indicate the subsections and the shaded region indicates
the absence of an accurate approximation. In Fig.~\ref{fig2}, we
depict typical profiles for the wave functions $\psi_{_1}$ and
$\psi_{_2}$ which are met in the four considered regimes. First of
all, in A, we focus on the case of weak segregation i.e.
$K\rightarrow 1$. Secondly in B, we develop an expansion around
the point of strong segregation $1/K\rightarrow 0$; the result is
expansion~\eqref{strongrepultsion} which is accurate for a broad
range of values. Thirdly, we treat the case of a strong healing
length asymmetry, that is, when $\xi_{_2}/\xi_{_1}\ll 1$, in C.
Finally, we study the case of $\xi_{_2}/\xi_{_1}\ll 1$ in
combination with strong segregation $1/K\rightarrow 0$ in D.

\subsection{Weak Segregation}\label{sec_weaksegr}
For the regime of weak segregation i.e. close to the point where
the two phases tend to mix ($K\rightarrow 1$), it was found by
Barankov in Ref.~\onlinecite{barankov} that the interface tension
varies as:
\begin{align}\label{weaksegregat}
\gamma_{_{12}}=4P\xi_{_1}\frac{\sqrt{K-1}}{3}\left(\frac{1-[\xi_{_2}/\xi_{_1}]^3}{1-[\xi_{_2}/\xi_{_1}]^2}\right).
\end{align}
In case of $\xi_{_2}/\xi_{_1}=1$, the same leading behavior was
recovered in Refs.~\onlinecite{mazets},~\onlinecite{malomed}
and~\onlinecite{ao} and the square-root behavior as a function of
$K-1$ arises from simple models as given in
Refs.~\onlinecite{timmermans} and~\onlinecite{ao}.
Approximation~\eqref{weaksegregat} is exact at $K=1$ but is
already $3.6\%$ off at $K=1.1$ when $\xi_{_2}=\xi_{_1}$. Despite
this small range of validity, such small values for $K-1$ were
realized in the early experiments on phase-segregated $^{87}$Rb
mixtures~\cite{stenger,hall,miesner,stamper}. On the other hand,
the use of an interface tension for determining the excess energy
in these experiments is inaccurate as the ``interface thickness'',
that is $\xi_{_1}/\sqrt{K-1}$, is of the same order as the length
of variation of the trapping potential; this invalidates the
assumption of constant chemical potentials across the interface.

\subsection{Strong Segregation}\label{sec_strongseg}
In the following we develop an expansion around the point of
strong segregation where $1/K\rightarrow 0$. An analogous
expansion which originated in a paper by Ginzburg and
Landau~\cite{ginzburg}, was used to determine an accurate
analytical expression for the interface tension of a
normal-superconducting interface~\cite{boulter}.

\begin{figure}
\begin{center}
   \epsfig{figure=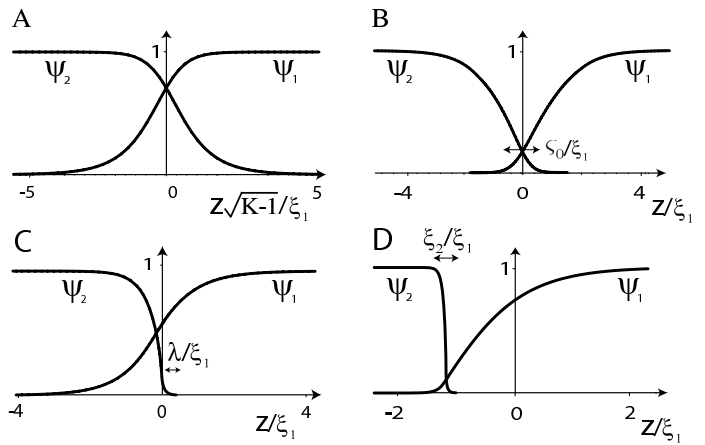,angle=0,width=250pt}
        \caption{Typical interface profiles
        for the four cases which are treated in
        Sects.~\ref{sec_inttens}A-D. We indicate the characteristic length scales over which the wave functions vary. A) Weak segregation regime
        $K\rightarrow 1$ (here with $\xi_{_2}/\xi_{_1}=1$). The relevant length scale is $\xi_{_1}/\sqrt{K-1}$
        which is large. B) Strong segregation regime with
        $1/K\rightarrow 0$ (here with $K=200$ and $\xi_{_2}/\xi_{_1}=1$). The condensates overlap over a
        length scale $\varsigma_{_0}$ (see Eq.~\eqref{varsigma}). C) The regime of strong
        healing length asymmetry $\xi_{_2}/\xi_{_1}\rightarrow 0$.
        When $\xi_{_2}/\xi_{_1}$ is small but nonzero, $\psi_{_2}$
        vanishes with a tail of length $\lambda$ (see Eq.~\eqref{lambda}). D) When both $\xi_{_2}/\xi_{_1}\rightarrow
        0$ and $1/K\rightarrow 0$, the condensates will overlap
        over a distance $\xi_{_2}$.
        \label{fig2}
        }
  \end{center}
\end{figure}

In the limit $1/K\rightarrow 0$, species $1$ and $2$ do not
overlap. Their wave functions are easily found using the GP
Eqs.~\eqref{GP} and the boundary conditions~\eqref{boundconds}:
$\psi_{_1}=\Theta(z)\tanh[z/\sqrt{2}\xi_{_1}]$ and
$\psi_{_2}=\Theta(-z)\tanh[-z/\sqrt{2}\xi_{_2}]$, where $\Theta$
is the Heaviside function. For a finite but small value of $1/K$,
these $\tanh$ profiles shift towards each other so as to overlap
over a length proportional to
$\sqrt{\xi_{_1}\xi_{_2}}/\sqrt[4]{K}$. In this overlap zone, the
wave functions are modified due to the interspecies interactions.
In Fig.~\ref{fig2}B, we depict such wave functions for the case of
$\xi_{_2}/\xi_{_1}=1$. Note that the length of overlap vanishes
very slowly when $1/K\rightarrow 0$ as it is proportional to
$1/\sqrt[4]{K}$. We introduce the following (change of) variables
($i=1,2$):
\begin{align}\label{varsigma}
& \widehat{z}\equiv z/\varsigma_{_0},\quad\quad
\varsigma_{_0}\equiv\frac{\sqrt{\xi_{_1}\xi_{_2}}}{\sqrt[4]{K}}\quad\text{and}\quad
\varsigma_{_i}\equiv\varsigma_{_0}/\xi_{_i}.
\end{align}
Note that $\varsigma_{_0}$ has the dimension of length (drawn in
Fig.~\ref{fig2}B) whereas $\varsigma_{_1}$ and $\varsigma_{_2}$
are dimensionless. We expand the wave function of phase $i=1,2$ in
terms of the parameter $\varsigma_{_i}$\label{varsigmadef}:
\begin{align*}
\psi_{_i}=\psi_{_i}^{_0}+\varsigma_{_i}(\psi_{_{i1}}-\psi_{_{i1}}^{_0})
+\varsigma_{_i}^3(\psi_{_{i2}}-\psi_{_{i2}}^{_0})+\ldots.
\end{align*}
We remark that $(\psi_{_i}-\psi_{_i}^{_0})/\varsigma_{_i}$ is a
regular function of $\varsigma_{_i}^2$. The functions
$\psi_{_{i1}}^{_0}$ and $\psi_{_{i2}}^{_0}$ are derivable from the
asymptotic behavior of $\psi_{_i}$ for $|\widehat{z}|\rightarrow
\infty$ which are shifted $\tanh$ profiles:
\begin{subequations}\label{profileexpansions}
\begin{align}
\psi_{_{1}}^{_0}&=\Theta(\widehat{z}+\delta_{_{11}}
+\varsigma_{_1}^2\delta_{_{12}}+\ldots)\nonumber\\
&\quad\quad\times\tanh\left[\frac{\varsigma_{_1}(\widehat{z}+\delta_{_{11}}
+\varsigma_{_1}^2\delta_{_{12}}+\ldots)}{\sqrt{2}}\right]\nonumber\\
&\equiv \varsigma_{_1}\psi_{_{11}}^{_0}
+\varsigma_{_1}^3\psi_{_{12}}^{_0}+\ldots\\
\psi_{_{2}}^{_0}&=\Theta(-\widehat{z}+\delta_{_{21}}
+\varsigma_{_2}^2\delta_{_{22}}+\ldots)\nonumber\\
&\quad\quad\times\tanh\left[\frac{\varsigma_{_2}(-\widehat{z}+\delta_{_{21}}
+\varsigma_{_2}^2\delta_{_{22}}+\ldots)}{\sqrt{2}}\right]\nonumber\\
&\equiv \varsigma_{_2} \psi_{_{21}}^{_0}
+\varsigma_{_2}^3\psi_{_{22}}^{_0}+\ldots
\end{align}
\end{subequations}
From this expansion, it is clear that $\psi_{_{i1}}^{_0}$ and
$\psi_{_{i2}}^{_0}$ are first-order and third-order polynomials in
$\widehat{z}$ respectively. Note that the ``absolute shift'' can
be set to zero at every order without any loss of generality; thus
$\delta_{_{11}}=\delta_{_{21}}$ and
$\varsigma_{_1}^2\delta_{_{12}}=\varsigma_{_2}^2\delta_{_{22}}$~\cite{boulter}.

 Substitution of the expanded wave
functions in the GP Eqs.~\eqref{GP} entails four new differential
equations:
\begin{subequations}\label{landauequations}
\begin{align}
\ddot{\psi}_{_{11}}&=\psi_{_{11}}\psi_{_{21}}^{2},\\
\ddot{\psi}_{_{21}}&=\psi_{_{21}}\psi_{_{11}}^{2},\\
\ddot{\psi}_{_{12}}&= -\psi_{_{11}}+
\psi_{_{12}}\psi_{_{21}}^{2}+2\psi_{_{11}}\psi_{_{21}}
\psi_{_{22}}[\xi_{_2}/\xi_{_1}]^{-2},\label{landauequationsc}\\
\ddot{\psi}_{_{22}}&= -\psi_{_{21}}+
\psi_{_{22}}\psi_{_{11}}^{2}+2\psi_{_{21}}\psi_{_{11}}\psi
_{_{12}}[\xi_{_2}/\xi_{_1}]^2.\label{landauequationsd}
\end{align}
\end{subequations}
Here the dot denotes the derivative with respect to $\widehat{z}$.
The solutions of these equations will provide us with the
numerical values for the $\delta$'s; indeed, the boundary
conditions for $\psi_{_{i1}}$ and $\psi_{_{i2}}$ are to
tangentially approach $\psi_{_{i1}}^{_0}$ and $\psi_{_{i2}}^{_0}$
respectively for $\widehat{z}\rightarrow\pm\infty$, and to vanish
for $\widehat{z}\rightarrow\mp\infty$. Now, in the same fashion as
the wave functions, one can expand the interface tension:
\begin{align*}
\gamma_{_{12}} &=\gamma^{_0}+\varsigma_{_0}\sum_{i=1,\, 2}
\left[\gamma_{_{i1}}-\gamma^{_0}_{_{i1}}+\varsigma_{_i}^2(\gamma_{_{i2}}-\gamma^{_0}_{_{i2}})+\mathcal{O}(\varsigma_{_i}^4)\right].
\end{align*}
The zeroth-order term is obtained using (unshifted) $\tanh$
profiles such that
$\gamma^{_0}=4\sqrt{2}P\left(\xi_{_1}+\xi_{_2}\right)/3$~\cite{ao,barankov},
while:
\begin{align*}
\sum_{i=1,\, 2}\gamma_{_{i1}}&=4P\int_{_{-\infty}}^
{_{\infty}}\text{d}\widehat{z}\,
\left(\dot{\psi}_{_{11}}^{2}+\dot{\psi}_{_{21}}^{2}\right),\\
\sum_{i=1,\,
2}\varsigma_{_i}^2\gamma_{_{i2}}&=\frac{8P}{\sqrt{K}}\int_{_{-\infty}}^{_{\infty}}\text{d}
\widehat{z}\,\left(
[\xi_{_2}/\xi_{_1}]\dot{\psi}_{_{11}}\dot{\psi}_{_{12}}+\frac{\dot{\psi}_{_{21}}\dot
{\psi}_{_{22}}}{[\xi_{_2}/\xi_{_1}]}\right).
\end{align*}
After long calculations, one can express the interface tension in
terms of the spatial shifts $\delta_{_{11}}$, $\delta_{_{12}}$,
$\delta_{_{21}}$ and $\delta_{_{22}}$, in a way similar to what
was done in Ref.~\onlinecite{boulter}:
\begin{figure}
\begin{center}
   \epsfig{figure=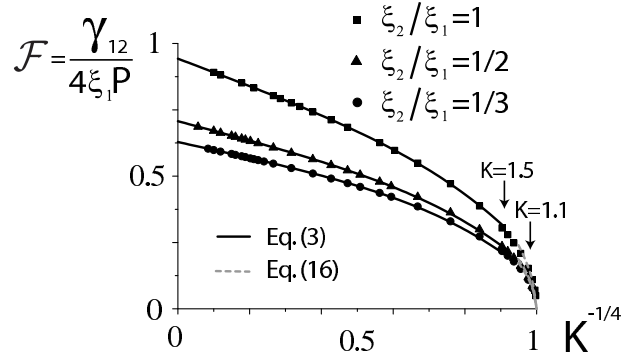,angle=0,width=230pt}
        \caption{Values of $\mathcal{F}$ or interface tension $\gamma_{_{12}}$ in units of $4P\xi_{_1}$,
        against values of $K^{-1/4}$ where $K=G_{_{12}}/\sqrt{G_{_{11}}G_{_{22}}}$. The squares,
        triangles and dots are obtained by numerical integration of the GP
        equations for $\xi_{_2}/\xi_{_1}=1,1/2$ and $1/3$
        respectively. The full lines are plots of
        Eq.~\eqref{strongrepultsion} and are seen to give an
        accurate approximation for values of K in the interval
        $[1.5,\,\infty[$. The grey lines plot
        Eq.~\eqref{weaksegregat} which is accurate for $K\in[1,1.1]$.
        \label{fig3}
        }
  \end{center}
\end{figure}
\begin{align*}
\gamma_{_{12}}=4P\xi_{_1}&\left[\frac{\sqrt{2}}{3}\left(1+\xi_{_2}/\xi_{_1}\right)-\frac{(\delta_{_{11}}+\delta_{_{21}})}{3}\frac{\sqrt{\xi_{_2}/\xi_{_1}}}{\sqrt[4]{K}}\right.\\
&\quad\left.-\frac{\sqrt{\xi_{_2}/\xi_{_1}}\left(\delta_{_{12}}[\xi_{_2}/\xi_{_1}]+\delta_{_{22}}[\xi_{_1}/\xi_{_2}]\right)}{\sqrt[4]{K}\sqrt{K}}+\ldots\right].
\end{align*}
It is readily seen in the expansion that $\gamma_{_{12}}$ is
symmetrical in $\xi_{_1}$ and $\xi_{_2}$, as it must be. As a
final step, we extracted numerical values for the $\delta$'s from
the numerical integration of Eqs.~\eqref{landauequations}. We find
that $\delta_{_{11}}=\delta_{_{21}}=0.771$~\cite{boulter}, and,
assuming $\xi_{_2}/\xi_{_1}=1$ in Eqs.~\eqref{landauequationsc}
and \eqref{landauequationsd}, we obtain
$\delta_{_{12}}=\delta_{_{22}}= 0.055$. Note that this result is
different from $\delta_{_2}$ found in Ref.~\onlinecite{boulter}
because the problem at hand is different starting from the first
order in $\varsigma_{_i}^2$. By fitting numerically-obtained
values for $\gamma_{_{12}}$ with an additional fitting term of
fifth order in $1/\sqrt[4]{K}$, and neglecting the dependence of
$\delta_{_{12}}$ and $\delta_{_{22}}$ on $\xi_{_2}/\xi_{_1}$, we
finally arrive at expression~\eqref{strongrepultsion}.

The accuracy of this expansion is clear from Fig.~\ref{fig3} where
we plot Eq.~\eqref{strongrepultsion} in full lines against the
numerically obtained values for $\xi_{_2}/\xi_{_1}=1$ (squares),
$\xi_{_2}/\xi_{_1}=1/2$ (triangles) and $\xi_{_2}/\xi_{_1}=1/3$
(dots). Our expansion works very well for $K$ in the interval
$[1.5,\, +\infty[$; moreover, it is good for all three values of
$\xi_{_2}/\xi_{_1}$, which a posteriori justifies the neglect of
the dependence of $\delta_{_{12}}$, $\delta_{_{22}}$ and the
fitted higher-order term on $\xi_{_2}/\xi_{_1}$. Also in
Fig.~\ref{fig3}, we draw with grey lines Barankov's
approximation~\eqref{weaksegregat}, accurate for values of $K$
between $1$ and $1.1$.

\subsection{Strong Healing Length Asymmetry}\label{sec_smallcoh} So far, we have
constructed expansions around extreme values for the interspecies
interaction parameter $K$. In the following, we will focus on the
case when the healing length of species $2$ is much smaller than
the one of species $1$ i.e. when $\xi_{_2}/\xi_{_1}\ll 1$.

In the limit of $\xi_{_2}/\xi_{_1}=0$ (and such that
$\xi_{_2}^2\ddot{\psi}_{_{2}}\rightarrow 0$), the GP
Eqs.~\eqref{GP} can be reduced to
\begin{subequations}
\begin{align}
\psi_{_{2}}^2=&\,\Theta(-z)\left[1-K\psi_{_{1}}^2\right],\label{kleinexia}\\
\xi_{_1}^2\ddot{\psi}_{_{1}}=&\,\Theta(z)\psi_{_{1}}\left[-1+\psi_{_{1}}^2\right]\nonumber\\
&+\Theta(-z)\psi_{_{1}}(K-1)\left[1-(1+K)\psi_{_{1}}^2\right],\label{kleinexib}
\end{align}
\end{subequations}
from which it follows that the density of phase $2$ is slaved to
the density of phase $1$. The solution to Eq.~\eqref{kleinexib}
is~\cite{alexandrov}:
\begin{align}\label{alexandrov}
\psi_{_{1}}&=\Theta(z)\tanh\left(\frac{z+z_{_1}}{\sqrt{2}\xi_{_1}}\right)\\
&\,\quad+\frac{\Theta(-z)\sqrt{2}}{\sqrt{K+1}}\left[\cosh\left(\frac{z+z_{_2}}{\xi_{_1}/\sqrt{K-1}}\right)\right]^{-1}.\nonumber
\end{align}
The constants $z_{_1}>0$ and $z_{_2}<0$ are determined by the
continuity of $\psi_{_{1}}$ at $z=0$ where $\psi_{_{1}}^2(0)=1/K$.
The wave function $\psi_{_{2}}$ is nonzero only when $z<0$ and
relates to $\psi_{_{1}}$ by Eq.~\eqref{kleinexia}. Near $z=0$ it
vanishes with the square-root behavior
$\psi_{_2}\propto\sqrt{-z/\xi_{_1}}$.

The first-order corrections due to a small and nonzero
$\xi_{_2}/\xi_{_1}$ affect only $\psi_{_2}$, near the origin $z=0$
where $\ddot{\psi}_{_2}$ is large due to the square-root behavior
of $\psi_{_2}$. These corrections will induce wave function
$\psi_{_2}$ to acquire a small tail of length $\lambda$ which is
depicted Fig.~\ref{fig2}C. We search now how GP Eq.~\eqref{GPb}
can be modified to appropriately describe such corrections. We
assume that $\psi_{_1}$ varies little on the length scale
$\lambda$; this allows to expand $K\psi_{_1}^2$ in Eq.~\eqref{GPb}
around $z=0$ as $K\psi_{_1}^2\approx 1+Fz/\xi_{_1}$ where
$F\equiv\sqrt{2/K}(K-1)$. Sufficiently far from the origin, when
$\lambda\ll-z\ll\xi_{_1}$, we want $\psi_{_2}$ to vary as
$\psi_{_2}\propto\sqrt{-z/\xi_{_1}}$; therefore, we may not throw
away the nonlinear term $\psi_{_2}^3$ in Eq.~\eqref{GPb}. By
appropriately scaling $\psi_{_2}$ and $z$, we find that, for
$|z|\ll\xi_{_1}$,
\begin{align}\label{kleinexic}
\ddot{\overline{\psi}}_{_2}&=\overline{\psi}_{_2}(\overline{z}+\overline{\psi}_{_2}^2),
\end{align}
where the dot denotes the derivative with respect to
$\overline{z}$ which we define as:
\begin{align}\label{lambda}
\overline{z}\equiv z/\lambda\quad\text{with}\quad\lambda\equiv
\xi_{_2}[F[\xi_{_2}/\xi_{_1}]]^{-1/3},
\end{align}
and where the new wave function $\overline{\psi}_{_2}\equiv
\psi_{_2}[F[\xi_{_2}/\xi_{_1}]]^{-1/3}$. For
$-\overline{z}\gg\lambda$, the term $\ddot{\overline{\psi}}_{_2}$
in Eq.~\eqref{kleinexic} may be neglected; indeed, doing so
results in the solution $\overline{\psi}_{_2}=
\sqrt{-\overline{z}}$. As expected this is in fact
$\psi_{_{2}}\propto -\sqrt{z/\xi_{_1}}$.

The differential equation which appears in Eq.~\eqref{kleinexic}
was encountered earlier by Lundh et al.~\cite{lundh} and Dalfovo
et al.~\cite{dalfovo} when studying the BEC order parameter at the
border of a large harmonic trap. The role of the trapping
potential is played here by the interactions with condensate $1$.

We briefly sketch how we calculated the interface tension; we
refer to Refs.~\onlinecite{dalfovo} and \onlinecite{lundh} for
details of the method. To zeroth order in $\xi_{_2}/\xi_{_1}$, the
interface tension is easily found by integration of
$\dot{\psi}_{_1}^2$ (see Eq.~\eqref{interfacetension}) with
$\psi_{_1}$ given in Eq.~\eqref{alexandrov}. For small and nonzero
$\xi_{_2}/\xi_{_1}$, extra energy contributions arise due to the
integral over $\dot{\psi}_{_2}^2$ (see
Eq.~\eqref{interfacetension}). For $-z\gg\lambda$, $\psi_{_2}$
must be taken from Eqs.~\eqref{kleinexia} and \eqref{alexandrov}
whereas for $|z|\ll\xi_{_1}$, $\psi_{_2}$ is the (numerical)
solution of Eq.~\eqref{kleinexic}. Since these two solutions for
$\psi_{_2}$ are equal in the region $\lambda\ll -z\ll\xi_{_1}$,
where $\psi_{_2}\propto\sqrt{-z/\xi_{_1}}$ the interface tension
is well-defined. All the extra energy contributions due to a
nonzero $\xi_{_2}/\xi_{_1}$ can be seen to be of order
$[\xi_{_2}/\xi_{_1}]^2$.

Calculations then lead to $\mathcal{F}$ to second order in
$\xi_{_2}/\xi_{_1}$:
\begin{widetext}
\begin{align}\label{resultkleinexi}
\frac{\gamma_{_{12}}}{4\xi_{_1}P}=&
\frac{2}{3}\frac{\sqrt{K-1}}{K+1}
\left[1-\left(\frac{K-1}{2K}\right)^{3/2}\right]
+\frac{\sqrt{2}}{3}\left(1+\frac{1}{2}\left(\frac{1}{K}\right)^{3/2}
-\frac{3}{2\sqrt{K}}\right)+[\xi_{_2}/\xi_{_1}]^2\frac{\sqrt{K-1}}{3}\left(\frac{K+3}{K+1}\right)\nonumber\\
&+[\xi_{_2}/\xi_{_1}]^2\frac{(K-1)}{\sqrt{2K}}\left[-\frac{2}{3}\frac{(K+2)}{(K+1)}
+0.7+\frac{\ln}{6}
\left(\frac{32K(K-1)}{(K+1)^3[\xi_{_2}/\xi_{_1}]^2}\right)
-\text{arctanh}\left(\sqrt{\frac{K-1}{2K}}\right)\right]+\ldots
\end{align}
\end{widetext}
Since there are two length scales ($\lambda$ and $\xi_{_1}$) over
which the density varies, their relation $\lambda\ll\xi_{_1}$
renders difficult the full numerical integration of the GP Eqs.
Accordingly, we could not verify expansion~\eqref{resultkleinexi}
directly.

Conditions for the validity of the above approximation can be
derived. First, the linearization of $\psi_{_1}$ near $z=0$ in
Eq.~\eqref{GPb} is justified only when the length over which
higher-order corrections to $\psi_{_1}$ are relevant (that is
$[\dot{\psi}_{_1}/\ddot{\psi}_{_1}](0)\propto\sqrt{K}\xi_{_1}$) is
large as compared to the length over which $\psi_{_2}$ varies
(being $\lambda$). Second, corrections to $\psi_{_1}$ caused by a
nonzero $\xi_{_2}/\xi_{_1}$ can be neglected on
$[-\lambda,\infty[$ when, in Eq.~\eqref{GPa}, the amplitude
$K\psi_{_2}^2$ of the term $K\psi_{_2}^2\psi_{_1}$ (being
$K[F[\xi_{_2}/\xi_{_1}]]^{2/3}$) is sufficiently small. These
conditions of validity amount to:
\begin{align*}
\xi_{_2}/\xi_{_1}\ll\sqrt{K(K-1)}\quad\text{and}\quad
\xi_{_2}/\xi_{_1}\ll \sqrt{K}.
\end{align*}
Thus, our expansion fails in the regime of strong segregation when
$\sqrt{K}\ll\xi_{_2}/\xi_{_1}$ and close to weak segregation when
$\sqrt{K-1}\ll \xi_{_2}/\xi_{_1}$. Note that to zeroth order in
$\xi_{_2}/\xi_{_1}$, our expansion~\eqref{resultkleinexi} agrees
with~\eqref{weaksegregat} when $K\rightarrow 1$. However, it
differs from the expansion~\eqref{weaksegregat} by a factor of $2$
in the term of order $[\xi_{_2}/\xi_{_1}]^2$.

\subsection{Both Strong Segregation and Strong Healing Length Asymmetry}\label{sec_both}
One may now wonder what happens to the interface tension when both
$K$ and $[\xi_{_2}/\xi_{_1}]^{-1}$ are large. When $K$ goes faster
to infinity than $[\xi_{_2}/\xi_{_1}]^{-2}$, we will find that
expansion~\eqref{strongrepultsion} is still valid while
expansion~\eqref{resultkleinexi} must be taken in the inverse
case. In the following, we focus on the intermediate case when
both $[\xi_{_2}/\xi_{_1}]^{-2}$ and $K$ diverge such that
$\kappa\equiv([\xi_{_2}/\xi_{_1}]\sqrt{K})^{-1}$ remains of order
one. To zeroth order in both $\xi_{_2}/\xi_{_1}$ and $K^{-1}$, the
condensate wave functions do not overlap; seen on a lengthscale
$\xi_{_1}$, this results in the wave functions
$\psi_{_2}=\Theta(-z)$ and
$\psi_{_1}=\Theta(z)\tanh[z/\sqrt{2}\xi_{_1}]$.

For small and nonzero values for $\xi_{_2}/\xi_{_1}$ and $1/K$
while
$\kappa\equiv([\xi_{_2}/\xi_{_1}]\sqrt{K})^{-1}$\label{kappadef}
is of order one, the condensates overlap. Due to the strong
repulsion this happens only within a short region of length
$\xi_{_2}$; over that interval, the value of the wave function
$\psi_{_2}$ varies between zero and one while the value of
$\psi_{_1}$ does not vary much (see Fig.~\ref{fig2}D). The latter
is only modified close to $z=0$, where, to zeroth order, it
vanishes linearly. This brings about the definition of a new wave
function $\phi_{_1}$:
\begin{align*}
\psi_{_1}=[\xi_{_2}/\xi_{_1}]\left[\phi_{_1}-\frac{z'\Theta(z')}{\sqrt{2}}\right]+
\Theta(z)\tanh\left(\frac{z}{\sqrt{2}\xi_{_1}}\right).
\end{align*}
Here we introduced the dimensionless variable $z'\equiv
z/\xi_{_2}$ and $\phi_{_1}$ must have the asymptotic behavior
$\phi_{_1}(z'\rightarrow\infty)=z'/\sqrt{2}$ and
$\phi_{_1}(z'\rightarrow-\infty)=0$. We can now expand the GP
Eqs.~\eqref{GP} in terms of the small parameters
$[\xi_{_2}/\xi_{_1}]^2$ and $1/K$ in the overlap interval:
\begin{align}\label{ginzburg}
\frac{\ddot{\phi}_{_1}}{\phi_{_1}}=&\left(\frac{\psi_{_2}}{\kappa}\right)^{2}
\quad\text{and}\quad\frac{\ddot{\psi}_{_2}}{\psi_{_2}}=
-1+\psi_{_2}^{2} +\left(\frac{\phi_{_1}}{\kappa}\right)^{2}.
\end{align}
The dots denote the derivative with respect to $z'$. One may
notice that these equations are the same as the Ginzburg-Landau
equations which are valid in a superconductor when a constant
magnetic field is applied parallel to its surface. In that case,
the Ginzburg-Landau parameter $\kappa$ is the ratio of the
penetration length of the magnetic field to the coherence length,
$\psi_{_2}$ plays the role of the order parameter of the
superconductor and $\phi_{_1}$ that of the vector
potential~\footnote{See Ref.~\onlinecite{indekeu2} for more
details such as the choice of the gauge.}.

By a straightforward calculation, one can rewrite the interface
tension Eq.~\eqref{interfacetension} as an integral over
$\phi_{_1}$ and $\psi_{_2}$:
\begin{align}\label{kappaint}
\frac{\gamma_{_{12}}}{4\xi_{_1}P}
=&\frac{\sqrt{2}}{3}+[\xi_{_2}/\xi_{_1}]\int\text{d}z'\,\left[
\left(\dot{\phi}_{_1}-\frac{1}{\sqrt{2}}\right)^2+\dot{\psi}_
{_2}^{2}\right],
\end{align}
where the dots denote the derivative with respect to $z'$ and
$\phi_{_1}$ and $\psi_{_2}$ are obtained by the Euler-Lagrange
Eqs.~\eqref{ginzburg}.

An analytical expression for the interface tension of a
superconductor-normal interface in case of small $\kappa$ was
obtained by Boulter and Indekeu in Ref.~\onlinecite{boulter}.
Using their result, one finds:
\begin{align}\label{bothstrongrepandsmallcoherence}
\frac{\gamma_{_{12}}}{4P\xi_{_1}}=&\frac{\sqrt{2}}{3}+[\xi_{_2}/\xi_{_1}]\left(\frac{\sqrt{2}}{3}-\frac
{0.5140}{([\xi_{_2}/\xi_{_1}]\sqrt{K})^{1/2}}\right.\\
&\left.-\frac{0.06653}{([\xi_{_2}/\xi_{_1}]\sqrt{K})^{3/2}}+\frac
{0.00107}{([\xi_{_2}/\xi_{_1}]\sqrt{K})^{5/2}}+\ldots\right).\nonumber
\end{align}
This approximation is very accurate for values of $\kappa$ in the
interval between $0$ and $1$~\cite{boulter}. To first order in
$\kappa$, Eq.~\eqref{bothstrongrepandsmallcoherence} agrees with
the interface tension found for $K\rightarrow\infty$ as given in
Eq.~\eqref{strongrepultsion}.

 For large $\kappa$,
according to standard results, the integral in
Eq.~\eqref{kappaint} goes as $2(1-\sqrt{2})\kappa/3$ such
that~\cite{saintjames}:
\begin{align}\label{saintjamesexp}
\frac{\gamma_{_{12}}}{4P\xi_{_1}}=&\frac{\sqrt{2}}{3}+\frac{2(1-\sqrt{2})}{3\sqrt{K}}+[\xi_{_2}/\xi_{_1}]\mathcal{O}(\kappa^{-1}).
\end{align}
Since large $\kappa$ implies $\xi_{_2}/\xi_{_1}\ll\sqrt{K}$, this
last result is in full accord with the $K\rightarrow\infty$ limit
in expansion Eq.~\eqref{resultkleinexi}. The higher-order
corrections to Eq.~\eqref{saintjamesexp} can also be extracted
from Eq.~\eqref{resultkleinexi}.

Note that Eqs.~\eqref{ginzburg} and
\eqref{bothstrongrepandsmallcoherence} are valid for all values of
$\kappa$ and may be generally used to numerically integrate the
interface tension in the limit under consideration.

\section{Temperature Dependence of the Interface Tension}\label{sec_tempdep}
At low but nonzero temperature, interface waves will be excited as
thermal excitations. In the following section, we will consider
these and ignore quantum fluctuations. It is common to incorporate
the free energy contributions which arise from capillary
excitations into a new temperature-dependent interface tension
$\gamma_{_{12}}(T)$. Generally, when the interface waves have
wavenumbers $k$ and frequencies $\omega(k)$, one finds for a
three-dimensional system that~\cite{atkins,landau4}:
\begin{align*}
\gamma_{_{12}}(T)=\gamma_{_{12}}(0)+\frac{k_{_B}T}{2\pi}\int_{_0}^{_\infty}\ln\left(1-e^{-\hslash
\omega(k)/k_{_B}T}\right)k\text{d}k,
\end{align*}
where $\gamma_{_{12}}(0)$ is the interface tension at zero
temperature and the last term embodies a negative correction due
to a nonzero temperature. For a flat interface between two BECs,
one may obtain the capillary-wave spectrum from the time-dependent
GP Eqs. For sufficiently large wavelengths $k^{-1}\gg
\gamma_{_{12}}/P$, one then finds for the modes with in-phase
motion of both species~\cite{mazets,barankov}:
\begin{align}\label{spectr}
\omega(k)=k^{3/2}
\sqrt{\frac{\gamma_{_{12}}}{m_{_{1}}n_{_{1}}+m_{_{2}}n_{_{2}}}},
\end{align}
where $n_i$ is the bulk density of phase $i=1,\,2$. When the
temperature $T$ satisfies the condition
$k_{_B}T\ll\hslash^2/(2m_{_i}(\gamma_{_{12}}/P)^2)$, only modes
with wavenumbers satisfying Eq.~\eqref{spectr} are thermally
excited. The temperature-dependent interface tension then
becomes~\footnote{A complete treatment of the interface
thermodynamics would include first of all capillary excitations
which are coupled to ``bulk excitations''; e.g.~a phonon incoming
to the interface which (partially) excites an interface excitation
and is then reflected and transmitted. Such modes are included for
the free interface of $^4$He by Saam in Ref.~\onlinecite{saam}.
Secondly, a full treatment would include the excitations which
involve out-of-phase motions of the two BEC components on either
side of the interface. Such modes were obtained by Mazets in
Ref.~\onlinecite{mazets}. However, all these additional modes are
unimportant at low temperatures since they have a high energy at
low wavenumber by their dispersion relations $\omega\propto k$ or
$\omega\propto \sqrt{k}$~\cite{saam,mazets}.}:
\begin{align*}
\gamma_{_{12}}(T)=\gamma_{_{12}}(0)-\frac{c\left(k_{_B}T\right)^{7/3}}{4\pi
\hslash^{4/3}}\left(\frac{m_{_{1}}n_{_{1}}+m_{_{2}}n_{_{2}}}{\gamma_{_{12}}(0)}\right)^{2/3},
\end{align*}
where $c\equiv\Gamma(7/2) \zeta(7/2)$ and $\zeta$ and $\Gamma$ are
the zeta and gamma function, respectively.

Taking $n_{_{1}}\approx n_{_{2}}$, $\gamma_{_{12}}(0)\propto
\xi_{_1}P$ and $\mu_{_{1}}\approx \mu_{_{2}}$, one calculates that
the relative temperature corrections
$[\gamma_{_{12}}(T)-\gamma_{_{12}}(0)]/\gamma_{_{12}}(0)$ are of
order $\sqrt{n_{_{1}}a_{_{11}}^3}(k_{_B}T/\mu_{_1})^{7/3}$. It is
known that $\sqrt{n_{_{1}}a_{_{11}}^3}\ll 1$ is the condition for
the GP theory to be applicable; moreover for the current BEC
experiments $k_{_B}T$ is of the same order or less than
$\mu_{_1}$~\cite{pita}.

On the other hand, for a system in the weakly segregated regime,
the interface tension varies as $\gamma_{_{12}}(0)\propto
\xi_{_1}P\sqrt{K-1}$ and accordingly vanishes as $K\rightarrow 1$.
The relative temperature corrections are then of order
$(K-1)^{-5/6}\sqrt{n_{_{1}}a_{_{11}}^3}(k_{_B}T/\mu_{_1})^{7/3}$
which may become large as $K\rightarrow 1$.

In conclusion, thermal fluctuations of a flat BEC interface do not
affect the interface tension at low temperature, except possibly
near weak segregation. Note that the calculated temperature
dependence of the interface tension does not hold in trapped
systems, as the interface modes there are
quantized~\cite{svidzinsky,lazarides}.

\begin{figure}
\begin{center}
   \epsfig{figure=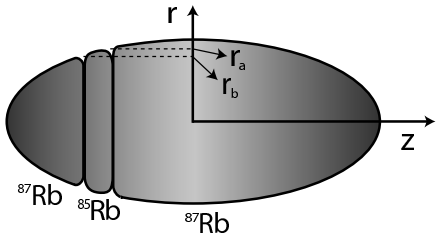,angle=0,width=150pt}
        \caption{Cross section of a trapped cloud of BEC mixtures as it was observed in Ref.~\onlinecite{papp2}. Due to phase segregation,
        $^{85}$Rb forms a layer between clouds of $^{87}$Rb. In
        Eq.~\eqref{harm}, we calculate the total excess energy
        due to the presence of the two interfaces.
        \label{fig4}
        }
  \end{center}
\end{figure}
\section{Discussion}\label{sec_exp}
We discuss now the relevance of our results with respect to the
recent experiments of Papp et al.~\cite{papp2}. Two main values
for the scattering lengths of phase-segregated states were
reported: one configuration was characterized by $K=3.01$ and
$\xi_{_2}/\xi_{_1}=0.84$ such that
$\mathcal{F}=\gamma_{_{12}}/(4P\xi_{_1})=0.434$, while for the
other configuration, $K=2.36$ and $\xi_{_2}/\xi_{_1}=0.952$ such
that $\mathcal{F}=0.415$. Here species $1$ and $2$ are $^{85}$Rb
and $^{87}$Rb respectively.  Note that these values for
$\mathcal{F}$ are obtained both numerically and with
Eq.~\eqref{strongrepultsion}. The values for $\mathcal{F}$
together with Eq.~\eqref{totalexcess} allow a numerical
calculation of the interface excess of any trapped configuration.

As an example, we consider now the topology as depicted in
Fig.~\ref{fig4} in which a layer of $^{85}$Rb BEC is present
between clouds of $^{87}$Rb. This configuration was experimentally
obtained in Ref.~\onlinecite{papp2} by strong radial confinement
so as to quench gravitational effects. Particles of phase $i=1,2$
are confined by an anisotropic trapping potential
$U_{_i}(r,z)=m_{_i}(\omega^{2}_{_{ri}}r^{2}+\omega^{2}_{_{zi}}z^{2})/2$
with $r^2=x^2+y^2$. Using Eq.~\eqref{totalexcess}, one finds the
total excess energy due to the presence of the two interfaces:
\begin{align}\label{harm}
\Omega_\mathcal{A}^{_{HO}}=\mathcal{F}(\xi_{_2}/\xi_{_1},K)\frac{m_{_1}^2\omega_{_{r1}}^3(r_{_a}^5+r_{_b}^5)}{20\hslash
a_{_{11}}}.
\end{align}
Here $r_{_a}$ and $r_{_b}$ are the maximal radial coordinates
along the two interfaces (see Fig.~\ref{fig4}) and
$\mathcal{F}=0.415$. Comparing this excess energy
$\Omega_\mathcal{A}^{_{HO}}$ with the total energy $E$ of a gas of
$1.4\times 10^{5}$ $^{87}$Rb particles, we obtain that
$\Omega_\mathcal{A}^{_{HO}}/E\approx 0.03$. This implies that,
even for the largest clouds observed in Ref.~\onlinecite{papp2},
the interface excess energy is substantial and must be considered
accurately when determining the shape of the ground state
configuration.

Note that in recent experiments on ultracold imbalanced fermion
gases, a breaking of the local density approximation was
encountered in the sense that the introduction of the interface
and its tension was essential in explaining the
experiments~\cite{partridge,vanschaeybroeck}.

The formalism presented in this work assumes a \textit{local
approximation} for the interface tension. For this approach to be
valid, the interface thickness must be smaller than the length
over which the trapping potential varies. Generally, the former
length is simply $\xi_{_1}+\xi_{_2}$; however, close to the mixed
state (i.e. when $K\rightarrow 1$) it is
$(\xi_{_1}+\xi_{_2})/\sqrt{K-1}$ which is larger. On the other
hand, the characteristic variation length of the trapping
potential of species $i$ is
$a_{_{ho}}=\sqrt{\hslash/m_{_i}\overline{\omega}_{_i}}$ with
$\overline{\omega}\equiv(\omega_{_{r,i}}^2\omega_{_{z,i}})^{1/3}$.
If we consider the configuration of $1.4\times 10^5$ particles
with $K=2.36$ and $\xi_{_2}/\xi_{_1}=0.952$, we find that
$(\xi_{_1}+\xi_{_2})/a_{_{ho}}\approx 0.1$ is indeed small at the
trap center~\cite{papp2}. Although the coherence lengths diverge
at the trap boundary, this divergence is so weak that
$\xi_{_1}+\xi_{_2}$ exceeds $a_{_{ho}}$ only in a layer of
thickness of $0.5\%$ of the trapping radius. This justifies the
use of the local approximation for the interface in the
experiments of Ref.~\onlinecite{papp2}.

\section{Conclusion}
We propose a method to calculate the total energy of a trapped
phase-segregated BEC mixture. In order to find the bulk energies,
one can use a local density approximation or Thomas-Fermi
approach. However, with respect to the recent experiments on
phase-segregated BEC components~\cite{papp2}, the bulk
contribution must be supplemented by interface excesses which may
have pronounced effects on the ground state topology. The
interface excess energy arises from the presence of the two-phase
boundaries, the position of which are determined by the Laplace
equation. Using a local approximation for the interface, we write
the total excess energy as a simple integral of the trapping
potential along the interface (see Eq.~\eqref{totalexcess}). By
use of analytical expansions, we find the interface tension in
almost all parameter regimes (see Fig.~\ref{fig1}), including the
regimes encountered in Ref.~\onlinecite{papp2}. We also study the
influence of thermal fluctuations on the interface tension at low
temperature and find that their effect is negligible, except
possibly close to the mixed state.

\section{Acknowledgements}
The author acknowledges partial support by Projects Nos.~FWO
G.0115.06 and GOA/2004/02 and thanks very much Joseph Indekeu for
useful discussions, suggestions and a thorough reading of the
manuscript. Achilleas Lazarides is thanked for corrections to the
manuscript. The author is supported by the Research Fund
K.U.Leuven.

\end{document}